\documentclass[12pt,preprint]{aastex}

\def\vec#1{\mbox{\boldmath $#1$}}

\slugcomment{Not to appear in Nonlearned J., 45.}

\shorttitle{Magnetic Structure Producing the flares in AR 11158}
\shortauthors{Inoue et al.}

\begin{document}


\title{Magnetic Structure Producing X- and M-Class Solar Flares 
       in Solar Active Region 11158}


\author{S.\ Inoue,}
\affil{School of Space Research, Kyung Hee University \\
       1, Seocheon-dong, Giheung-gu, Yongin, Gyeonggi-do 446-701,
       Republic of Korea}
\email{inosato@khu.ac.kr}

\author{K.\ Hayashi,}
\affil{W.\ W.\ Hansen Experimental Physics Laboratory, Stanford University \\
       Stanford, CA 94305, USA}

\author{D.\ Shiota}
\affil{Solar-Terrestrial Environment Laboratory, Nagoya University \\
       Furo-cho, Chikusa-ku, Nagoya, 464-8601, Japan}
\altaffiltext{1}{ Computational Astrophysics Laboratory, RIKEN(Institute 
       of Physics and Chemical Research),\\ Wako, Saitama 351-0198, Japan}
\altaffilmark{1}

\author{T.\ Magara}
\affil{School of Space Research,Kyung Hee University \\
       1, Seocheon-dong, Giheung-gu, Yongin, Gyeonggi-do 446-701,
       Republic of Korea}

\author{G.\ S.\  Choe}
\affil{School of Space Research,Kyung Hee University \\
       1, Seocheon-dong, Giheung-gu, Yongin, Gyeonggi-do 446-701,
       Republic of Korea}

  \begin{abstract}
   We study the three-dimensional magnetic structure of  solar active region 
   11158, which produced one X-class and several M-class flares on 2011 
   February 13$-$16. We focus on the magnetic twist in four flare events, 
   M6.6, X2.2, M1.0, and M1.1. The magnetic twist is estimated from the 
   nonlinear force-free field extrapolated from the vector fields obtained 
   from the Helioseismic and Magnetic Imager on board the Solar Dynamic 
   Observatory using magnetohydrodynamic relaxation method developed by 
   \cite{2011ApJ...738..161I}. We found that strongly twisted lines ranging
   from half-turn to one-turn twist were built up just before the M6.6- and 
   X2.2 flares and disappeared after that. Because most of the twist 
   remaining after these flares was less than half-turn twist, this result 
   suggests that the buildup of magnetic twist over the half-turn twist is 
   a key process in the production of large flares. On the other hand, even 
   though these strong twists were also built up just before the M1.0 and 
   M1.1 flares, most of them remained afterwords. Careful topological 
   analysis before the M1.0 and M1.1 flares shows that the strongly 
   twisted lines were surrounded mostly by the weakly twisted lines formed 
   in accordance with the clockwise motion of the positive sunspot, whose 
   footpoints are rooted in strong magnetic flux regions. These results imply 
   that these weakly twisted lines might suppress the activity of the strongly
   twisted lines in the last two M-class flares.
  \end{abstract}

  \section{Introduction}
   Solar active phenomena, observed as solar flares, coronal mass 
   ejection (CMEs), and filament eruptions, are driven by the release of  
   magnetic energy in the solar corona (\citealt{2002A&ARv..10..313P}). 
   Although many theoretical and numerical models of the magnetic field 
   dynamics have been proposed to date
   (e.g.,
    \citealt{2011LRSP....8....1C};
    \citealt{2011LRSP....8....6S}), 
   consensus has not yet been reached. On one hand, the optical 
   observations of the magnetic field using the Zeeman effect are limited on 
   the photosphere, therefore it is very difficult to understand the 
   complexity of the three-dimensional (3D) magnetic structure in the solar 
   corona as well as its physical properties and dynamics.

   On the other hand, the solar corona has been known to be in the low plasma 
   $\beta$ condition ($\beta$ $\approx$ 10$^{-2}$ $\sim$ 10$^{-1}$), which 
   means that the force-free condition is approximately satisfied 
   (\citealt{2001SoPh..203...71G}). Consequently, the nonlinear force-free 
   field (NLFFF) extrapolation becomes a an appropriate technique for 
   understanding the 3D magnetic structure 
   (\citealt{2006SoPh..235..161S};
    \citealt{2008SoPh..247..269M};
    \citealt{2012LRSP....9....5W}).
   The solar physics satellite {\it Hinode} (\citealt{2007SoPh..243....3K}) 
   can provide extremely high spatial resolution vector field data with the 
   Solar Optical Telescope (SOT; \citealt{2008SoPh..249..167T}). For example, 
   {\it Hinode} successfully observed the solar active region NOAA 10930, 
   which generated an X3.4 flare, and provide the vector field as time 
   series data covering before and after the flare. NLFFF extrapolation has 
   already been performed using these vector field data.
   \cite{2008ApJ...679.1629G} presented a 3D view of the core field composed  
   of sheared and twisted field lines lying above the polarity inversion line 
   (PIL), and \cite{2008ApJ...675.1637S} found that a strong current region 
   accumulated in the strongly sheared and twisted field lines before the 
   flare, most of which disappeared after the flare. 
   \cite{2008ApJ...676L..81J} clarified the energy variation in the altitude 
   direction during the flare. \cite{2011ApJ...738..161I}, 
   \cite{2012ApJ...747...65I} and \cite{2012ApJ...760...17I} also determined 
   the 3D NLFFF by using the magnetohydrodynamic (MHD) relaxation method and 
   the time series vector field data. They  quantitatively clarified the 
   variation in the twist profile of the magnetic field lines during the 
   flare, which ultimately leads the cause of the flare onset. Thus, the 
   NLFFF begins to clarify the 3D structure and physical properties such as 
   the stability, as well as the  evolution of the magnetic field, in the 
   solar active region; ultimately, it can even suggest the dynamics or onset 
   of the flare. Unfortunately, because the temporal resolution of these 
   vector field data is not sufficient for investigating the evolution of the 
   magnetic field, our understanding has not yet reached a phase that yields 
   a consensus.

   More recently, the Helioseismic and Magnetic Imager 
   (HMI; \citealt{2012SoPh..275..207S}; \citealt{Hoeksema2013} and see also http://hmi.stanford.edu/magnetic/) and 
   Atmospheric Imaging Assembly (AIA; \citealt{2012SoPh..275...17L}) on 
   board a new solar physics satellite, Solar Dynamic Observatory ({\it SDO}),
   can provide vector field data and extreme ultraviolet (EUV) images with 
   unprecedented temporal and spatial resolutions. Thus, we will have 
   many chances to analyze the 3D NLFFF with  higher accuracy in  space and 
   time. 

   On 2011 February 13$-$16, the solar active region NOAA 11158 produced one 
   X-class flare along with several M-class flares. This active region 
   consisted of bipoles showing strong shear and twist motion, and exhibited 
   coalescence of the same polarities and ultimately cancellation of opposite 
   polarities, along with complicated motion (\citealt{2012ApJ...744...50J}). 
   Fortunately, {\it SDO} continuously observed vector field data with a 
   cadence of 12 min and a spatial resolution of $\sim$ 0.5'' in a wide 
   region (216$\times$216 Mm$^{2}$), from February 12 to 16. Note that 
   the field of view of HMI/{\it SDO} is the full disk. Only the data in 
   the region, (216$\times$216 Mm$^{2}$), were released at that time by 
   HMI science team. The SOT/{\it Hinode} also provided high-resolution 
   vector field data whose field of view is smaller than that of 
   HMI/{\it SDO}; therefore, these data must help us to reconstruct the 3D  
   NLFFF with  high spatial and temporal resolution. 

   The profiles of the temporal evolution of the energy density, current 
   helicity or relative helicity estimated from the NLFFF 
   (\citealt{2012SoPh..tmp...67W}) were reported by 
   \cite{2012ApJ...752L...9J}. They found a bump structure in which the 
   magnetic helicity increased and decreased before the M6.6 and X2.2 flares, 
   whereas the magnetic energy and current helicity increased monotonically 
   before the X2.2 flare.
   \cite{2012ApJ...745L...4L} analyzed the vector field data related to the
   M6.6 flare that occurred on February 13 and found a specific region 
   close to the PIL in which a rapid increase in the horizontal field was 
   observed during the flare. They suggested that the 3D field line structure 
   obtained from the NLFFF has a shape favorable for tether-cutting 
   reconnection (\citealt{2001ApJ...552..833M}), which generates small 
   current-carrying loops close to the PIL that might be related to the 
   significant increase in the horizontal field. 
    \cite{2012ApJ...745L..17W} analyzed the magnetic field in terms of the 
   2D vector field and 3D NLFFF related to the X2.2 flare that occurred on 
   February 15. They also found a rapid enhancement of the horizontal 
   field close to the PIL on the photosphere and supported tether-cutting 
   reconnection as the origin of the flare.
    \cite{2012ApJ...748...77S} also extrapolated the 3D NLFFF based on  
   HMI/{\it SDO} data. They showed highly twisted field lines whose value 
   corresponds to 0.9 turn near the axis above the PIL, the profile along 
   the altitude of the magnetic energy, over 50 $\%$ of which is stored 
   below 6 Mm, eventually indicating that the numerically derived field 
   appeared to be more compact after the flare. Furthermore, 
   \cite{2012ApJ...757..149S} explained the non-radial eruption that occurred 
   to the east in terms of the magnetic topology of the field lines obtained 
   from the NLFFF.
   On the other hand, \cite{2012ApJ...757..147C} performed a data-driven 
   simulation in which an extrapolated NLFFF was driven by the normal 
   component of the magnetic field and the horizontal electric field derived 
   by solving the inverse problem of the induction equation 
   (\citealt{2010ApJ...715..242F}) from HMI/{\it SDO}. They successfully  
   reproduced the eruption in the X2.2 flare and, just after it showed a 
   rapid enhancement in the horizontal field on the photosphere. They 
   concluded that this enhancement resulted from relaxation of the arcade 
   field following a magnetic reconnection that produced a flux rope.

   Thus, solar active region NOAA 11158 is attractive for studies of  solar 
   flares. A wealth of vector field data and EUV images has been provided by 
   the {\it SDO} satellite. Nevertheless the conditions of the M- and X-class 
   flares and the differences in magnetic structure in these flare events are 
   not yet clear. The purpose of this study is to clarify them in terms of 
   the magnetic twist and topology obtained from the 3D NLFFF. The rest of 
   this paper is constructed as follows. The numerical method and 
   observational data set we used in this study are described in Section 2. 
   The results and discussion are presented in Sections 3 and 4,respectively. 
   Finally, the conclusion is summarized in Section 5. 

  \section{Numerical Method and Observations}  
  \subsection{NLFFF Extrapolation Based on the MHD-Relaxation Method}

   We numerically extrapolate the 3D coronal magnetic field assuming it as   
   the NLFFF by using the MHD relaxation method developed by 
  \cite{2011ApJ...738..161I},
  \cite{2012ApJ...747...65I} and
  \cite{2012ApJ...760...17I},
  which has already been applied to the solar active region NOAA 10930. 
  This method can numerically solve the following equations;

  \begin{equation}
  \frac{\partial \vec{v}}{\partial t} 
                        = - (\vec{v}\cdot\vec{\nabla})\vec{v}
                          + \frac{1}{\rho} \vec{J}\times\vec{B}
                          + \nu\vec{\nabla}^{2}\vec{v}.
  \end{equation}

  \begin{equation}
  \frac{\partial \vec{B}}{\partial t} 
                        =  \vec{\nabla}\times(\vec{v}\times\vec{B}
                        -  \eta\vec{J})
                        -  \vec{\nabla}\phi, 
  \label{induc_eq}
  \end{equation}

  \begin{equation}
  \vec{J} = \vec{\nabla}\times\vec{B}.
  \end{equation}
  
  \begin{equation}
  \frac{\partial \phi}{\partial t} + c^2_{h}\vec{\nabla}\cdot\vec{B} 
    = -\frac{c^2_{h}}{c^2_{p}}\phi,
  \label{div_eq}
  \end{equation}
  where $\vec{B}$ is the magnetic flux density, $\vec{v}$ is the velocity, 
  $\vec{J}$ is the electric current density, $\rho$ is the pseudo density, 
  and $\phi$ is a scalar potential. The pseudo density is assumed to be 
  proportional to $|\vec{B}|$ in order to ease the relaxation by equalizing 
  the Alfven speed in space. The last equation (\ref{div_eq}) is used to 
  avoid deviation from $\nabla\cdot\vec{B}=0$ and was introduced by 
  \cite{2002JCoPh.175..645D}. 
  
   The length, magnetic field, pseudo density, velocity, time, and electric 
  current density are normalized by   
  $L_0$ = 184.32 Mm,  
  $B_0$ = 2500 G, 
  $\rho_{0}$ = $|B_0|$,
  $V_{A}\equiv B_{0}/(\mu_{0}\rho_{0})^{1/2}$,    
  where $\mu_0$ is the magnetic permeability, whose value corresponds to 
   4$\pi \times 10^{-7}$ H/m,
  $\tau_{A}\equiv L_{0}/V_{A}$ and     
  $J_0=B_{0}/\mu_{0} L_{0}$,  
  respectively. The non-dimensional viscosity $\nu$ is set as a constant  
  $(1.0\times 10^{-3})$, and the non-dimensional resistivity $\eta$ is given 
  as 
  \begin{equation}
  \eta = \eta_0 + \eta_1 \frac{|\vec{J}\times\vec{B}||\vec{v}|^2}{\vec{|B|}},
  \end{equation} 
  where $\eta_0$ is a constant specific to each case whose values are shown 
  in Table \ref{tbl-1}, and $\eta_1=1.0\times 10^{-3}$ in non-dimensional 
  units. The second term is introduced to accelerate the relaxation to the 
  force-free field, particularly in the weak field region. The other 
  parameters $c_h^2$ and $c_p^2$ are set to the constants 0.04 and 0.1, 
  respectively.   

  The velocity field is capped at $v_{max}$ so that it does not correspond 
  to a large value. Let us define $v^{*} = |\vec{v}|/|\vec{v}_{A}|$ 
  and specify that if the value of $v^{*}$ becomes larger than the 
  value of $v_{max}$ given in Table \ref{tbl-1}, the velocity is modified as 
  follows;
  \begin{equation}
  \vec{v} \Rightarrow \frac{v_{max}}{v^{*}} \vec{v}.
  \end{equation}

  An initial condition is given by the potential field extrapolated using the  
  Fourier method from the normal component on the vector field assuming a 
  periodic condition and an exponential decay in the height direction, whose 
  analytical formula can be found in Equation (13.3.4) in 
  \cite{1994plph.book.....S}. First, we calculate the 3D potential field 
  according to the following equation,
  \begin{equation}
  \vec{B} = \Sigma \vec{b}_{(m,n)}^{\ast} exp(ik_m x + ik_n y - |k|z),
  \label{pote}
  \end{equation}
  where $k_m = 2\pi m/L_x$, $k_n = 2\pi n/L_y$, $|k| = \sqrt{k_m^2+k_n^2}$,  
  $m = -n_x/2, \cdots ,-1,1, \cdots, n_x/2$  and
  $n = -n_y/2, \cdots ,-1,1, \cdots, n_y/2$, respectively. 
  $\vec{b}_{(m,n)}^{\ast}$ 
  consists  of three components, 
  $b_{x(m,n)}^{\ast} = -i k_m b_{z(m,n)}^{\ast}/|k|$, 
  $b_{y(m,n)}^{\ast} = -i k_n b_{z(m,n)}^{\ast}/|k|$, 
  and $b_{z(m,n)}^{\ast}$. 
  However, after this method, vector field is slightly deviated from the 
  original one because $\vec{b}_{(0,0)}^{\ast}$ (averaged component) 
  was not counted in the process of the inverse Fourier transformation. 
  Therefore, as next step,  we calculate the potential field again using the 
  original vector field without removing the $\vec{b}_{(0,0)}^{\ast}$ and 
  the lateral and top boundaries of the 3D magnetic field obtained from 
  equation (\ref{pote}). The potential field is obtained from the normal 
  component of the magnetic field at all of the boundaries according to the 
  $\vec{\nabla}^2 \Phi=0$ (where $\vec{B}= -\vec{\nabla}\Phi$) after total 
  flux conservation $\int B_ndS=0$ is satisfied in the entire domain, where 
  $dS$ represents the surface element on all boundaries, and subscript $n$ 
  represents the component normal to the surfaces of the boundaries.

  The observed vector field is given as the bottom boundary condition, and 
  the three components of the magnetic field are fixed. However, the 
  observational data conditions cannot perfectly satisfy the induction 
  equation at the end, so the inconsistent boundary condition generates  
  errors related to $\vec{\nabla}\cdot\vec{B}$ near the boundary area. 
  Equation (\ref{div_eq}) can reduce these errors dramatically, as shown in 
  Table.1 (see D$=\int|\vec{\nabla}\cdot\vec{B}|^2dV$). On the top and side 
  boundaries, the transverse fields are determined by an induction equation 
  (\ref{induc_eq}) where a perfect conductive wall is assumed ( i.e., the 
  velocity and electric fields are set to zero), while the normal component 
  is kept with at the initial condition to conserve the total flux. Thus, 
  the side and top boundaries deviate from the real situation; therefore, 
  our analysis is limited to closed loops in the central area.

  The Runge-Kutta scheme with fourth-order accuracy for the temporal integral 
  and the central finite difference with second-order accuracy for the 
  spatial derivative are applied as the numerical scheme for this calculation.
  The numerical domain is covered by 184.32 $\times$ 184.32 
  $\times$ 184.32 ${\rm Mm^3}$, whose area is given by a 128 $\times$ 128 
  $\times$ 128 grid. The vector field given as the boundary condition 
  is obtained by 4 $\times$ 4 binning from the original vector field 512 
  $\times$ 512. The grid interval corresponds to about 1688 km/pixel i.e., 
  about 2.3''; therefore, we focus on the scale of the field lines connecting 
  the sunspots in the central area of the numerical domain. 
%
  
  \subsection{Observation of AR 11158}
  We analyzed one X-class flare (X2.2) that occurred at around 01:48 UT on 
  2011 February 15 and M-class flares (M6.6, M1.0, and M1.1) that occurred 
  at around 17:34 UT on February 13, and 01:36 UT and 07:40 UT on February 
  16, respectively. The GOES X-ray profile is shown in Figure \ref{f1}(a). 
  The observation of the vector field by HMI/{\it SDO}  covered this time 
  span, and HMI team has already provided these data, which are projected 
  and remapped to heliographic coordinates with a spatial resolution of 
  $\sim$ 0.5'' and a cadence of 12 min 
  (http://jsoc.stanford.edu/jsocwiki/ReleaseNotes). The vector magnetic 
  fields are obtained from the Very Fast Inversion of the Stokes Vector 
  algorithm (\citealt{2011SoPh..273..267B}) based on the Milne-Eddington 
  approximation, and the minimum energy method (\citealt{1994SoPh..155..235M};
  \citealt{2009SoPh..260...83L}) was used to solve the 180$^{\circ}$ ambiguity
  in the azimuth angle of the magnetic field. The vector fields before each 
  of the four flares are investigated (M6.6, X2.2, M1.0, and M1.1) as shown 
  in Figure \ref{f1}(b), in which the gray scale represents the distribution 
  of the normal component of the magnetic field ($B_z$).

  We use the {\ion{Ca}{2}} H 3968 {\AA} images in these flares for comparison
  with the extrapolated fields. These images were provided by the Broadband 
  Filter Imager on SOT on board {\it Hinode}. The fields of view are 
  188.28" $\times$ 111.58" with a pixel size of 1728 $\times$ 1024 at 
  17:35:38 UT on February 13, 111.58" $\times$ 111.58" with 
  1024 $\times$ 1024 at 01:50:18 UT on February 15, 
  188.28'' $\times$ 111.58'' with 1728 $\times$ 1024 at 01:40:39 on 
  February 16 and 111.58'' $\times$ 111.58'' with 1024 $\times$ 1024 at 
  07:42:13 on February 16. We also used the EUV images at 94 {\AA} at 
  23:59:28 UT on 2011 February 14 obtained from AIA/{\it SDO}, whose 
  view extracted from the full disk image corresponds to the same as that of 
  HMI used in this study. 

  \section{Results}

  \subsection{Reliability of 3D Magnetic Structure in AR 11158}
   First, we check the force-freeness of the reconstructed field. As an 
   example, Figure \ref{f2}(a) shows the force-free $\alpha$ map related 
   to the extrapolated field at 00:00 UT on February 15. The vertical and 
   horizontal axes represent the values of the force-free $\alpha$ estimated 
   on opposite footpoints of each field line. Note that these values 
   are estimated in the plane at 1688 km, i.e., 1 grid size in this case, 
   above the photosphere, and the field lines are traced from the region 
   where $B_z<-$50 G. Although the extrapolated field cannot satisfy the 
   force-free condition perfectly because of the deviation from the 
   force-free state in the photosphere, these dotted points seem to be 
   generally distributed along the green line, as shown in our previous 
   result (\citealt{2012ApJ...747...65I}).

   Next, we compare the 3D field lines structure with the observational 
   data at the various wavelengths to check its reliability. As an example, 
   Figure \ref{f2}(b) shows selected 3D  magnetic field lines in blue; they 
   are extrapolated from the vector field obtained from HMI/{\it SDO} at 
   00:00 UT on February 15, around 1.5 hr before the X2.2 flare, and plotted 
   over the distribution of the $B_z$ component. These selected field lines 
   start from points on the photosphere with $B_z >$250G. A strong shear 
   field clearly formed on the PIL between the sunspots located in the center 
   of the plot. This feature is consistent with earlier works (e.g., 
   \citealt{2012ApJ...748...77S}; \citealt{2012SoPh..tmp...67W};
   \citealt{2013arXiv1304.2979J}. 

    The same selected field lines are plotted with blue and gray lines 
   in Figure \ref{f2}(c) over the distribution of the $B_z$ component, and 
   then the brightest part of the AIA's 94 {\AA} image ($I$ $>$ 1.0$\times 
   10^{5}$(DN)) is superimposed. The blue lines are arbitrary chosen from 
   the field lines in Figure \ref{f2}(b) that locate in the strongly enhanced 
   areas of the AIA image. The inset shows the same image without the field 
   lines revealing high-intensity area in the EUV image is not only enhanced 
   near the central area but also extends in the east-west direction; 
   nevertheless, the blue lines can cover these regions. 
   
   Figure \ref{f2}(d) shows the another set of selected field lines in orange 
   and gray plotted over the $B_z$ distribution with the {\ion{Ca}{2}} images 
   obtained from FG/{\it Hinode}. Only the region with an {\ion{Ca}{2}} signal
   greater than 1000(DN) observed at 01:50:18 UT on February 15 is drawn in 
   this plot. The inset also shows the same picture without the field lines. 
   The orange field lines correspond to the strong shear field along the PIL 
   with respect to the gray lines, and their footpoints  are rooted in the 
   region where the {\ion{Ca}{2}} images are strongly enhanced. 

   This result is consistent with our previous study 
   (\citealt{2011ApJ...738..161I}), which confirms the reliability of the 
   extrapolated field as well as the fact that magnetic reconnection is 
   occurred along these strong non-potential field lines because the 
   strong {\ion{Ca}{2}} illumination is deeply related to magnetic 
   reconnection (\citealt{2002A&ARv..10..313P}; 
   \citealt{2000JGR...10523153F}).   

  \subsection{Twist Analysis of the Extrapolated Field in AR 11158}
  \subsubsection{Comparing Magnetic Twist with {\ion{Ca}{2}} Intensity 
              and Distribution}
   We estimate the magnetic twist from the 3D extrapolated field lines and 
   compare it with the {\ion{Ca}{2}} images obtained from SOT/{\it Hinode} 
   in detail. The magnetic twist is an important proxy for determining the 
   stability of the magnetic configuration 
   (\citealt{1958PhFl....1..265K};
    \citealt{1979SoPh...64..303H};
    \citealt{2004A&A...413L..27T};
    \citealt{2006ApJ...645..742I})
   as well as a convenient one for 
   clarifying its topology. The twist index is defined as
   \begin{equation}
   T_n = \int\frac{dT_n}{dl}dl = \int\frac{J_{||}}{4 \pi B_{||}}dl = 
         \frac{1}{4\pi} \int \alpha dl, 
   \label{eq_twist}
   \end{equation} 
   where the line integral $\int dl$ is taken along each magnetic field 
   line,  and the force-free $\alpha$ can be obtained from 
   $\alpha = \vec{J} \cdot \vec{B}/|\vec{B}|^2$ for each field line. The 
   physical significance is that it represents the degree of twist of a 
   magnetic field line corresponding to the measurement of the magnetic 
   helicity generated by the current parallel to a field line
   (\citealt{1984JFM...147..133B}; \citealt{2006JPhA...39.8321B};
    \citealt{2010A&A...516A..49T}; \citealt{2012ApJ...760...17I}). 
   Note that following the work of \cite{2006JPhA...39.8321B}, we applied  
   this definition only to closed loops connecting the sunspots, as shown 
   in Figure \ref{f3}.

   The red lines in the upper panels in Figures \ref{f3} represent the 
   contours of half-turn twist ($T_n$ = 0.5); i.e., the regions surrounded by 
   the red contours are dominated by strongly twisted lines ($T_n > 0.5$), 
   with the {\ion{Ca}{2}} images around the peak time of the GOES X-ray flux 
   in each flare. Hereafter, 'strongly twisted lines' will refer to field 
   lines having more than a half-turn twist ($T_n > $0.5). The white lines 
   represent contours of $|B_z|=$625 G. The M6.6 and X2.2 flares on 
   February 13 and February 15, respectively, reveal that the locations of 
   strongly twisted regions surrounded by the red lines correspond to enhanced
   regions of strong {\ion{Ca}{2}} illumination as shown in 
   Figure \ref{f2}(d). In particular, the profile of the strongly twisted 
   region ($T_n > 0.5$) at 00:00 UT on February 15 is similar to that of the 
   {\ion{Ca}{2}} image, indicating that the dramatic magnetic reconnection 
   occurred in the strongly twisted magnetic lines having around or more than 
   half-turn twist, as shown in \cite{2011ApJ...738..161I} and 
   \cite{2012ApJ...760...17I}. In addition, because the strongly twisted 
   regions on February 15 seem to be larger than that of February 13, this
   change in the size of the strongly twisted area might be related to the 
   occurrence conditions of X or M-class flares.

   On the other hand, the enhancement of the {\ion{Ca}{2}} image in the 
   M1.0 flare on February 16 appears outside the strongly twisted regions. 
   Therefore, the strongly twisted lines seem not to be related to this flare 
   despite the growth in the size of the twisted regions due to the clockwise 
   motion of the positive sunspot. However, in the M1.1 flare, the same 
   regions with strongly twisted lines coincided well with the areas of 
   enhanced {\ion{Ca}{2}}, whose values are smaller than that in the M6.6 or 
   X2.2 flares. At this time, the strongly twisted regions surrounded by the 
   red contours also seem to grow continuously.  
 
   The lower panels in Figures \ref{f3} show the selected field lines in 
   orange and blue corresponding to magnetic twist of more and less than a 
   half-turn twist, plotted over the upper panels. The footpoints of these 
   field lines lie in the regions where the {\ion{Ca}{2}} image is strongly 
   enhanced; i.e., these field lines connect the two-ribbon flare across the 
   PIL except that in M1.0 flare and weakly twisted lines in M1.1 flare. 
   According to these results, the magnetic field lines on February 15 are 
   obviously elongated compared with those on February 13 and 16, meaning 
   that stronger shear is formed just before the X2.2 flare compared to the M 
   flares. 

  \subsubsection{Temporal Evolution of the Magnetic Twist during the 
                 Major Flares}
   Figure \ref{f4} shows the temporal evolution of the magnetic twist. At 
   16:00 UT on February 13, the strongly twisted regions in the M6.6 flare, 
   which have  0.5 $< T_n <$ 1.0, are already built up by the clockwise motion
   of the positive sunspot. Most of the strongly twisted lines disappear from  
   the data after this flare. At 00:00 UT on February 15, before the X2.2 
   flare, the strongly twisted regions are built up again and seem to dominate
   larger areas than before the previous M6.6 flare. In contrast to the 
   previous flare, some parts of the strongly twisted regions remained even 
   after the X2.2 flare. Nevertheless, the store-and-release scenario of 
   magnetic helicity appeared clearly during the large flares. As a common 
   feature before each flare, we have never found a strongest twisted regions 
   having a value greater than one-turn twist ($T_n > $ 1.0). This means that 
   the magnetic configuration is stable against the ideal MHD instability; 
   however, even these magnetic configurations of less than one-turn twist 
   can generate large flares. 
  
    The twist profiles at 00:00 UT and 07:00 UT on February 16 are the 40-90 
   min before the M1.0 and M1.1 flares, respectively. The strongly twisted 
   regions of more than half-turn twist ($T_n$ = 0.5) are built up again 
   owing to the continuous clockwise twist and shear motions of the positive 
   sunspot. In contrast to the previous large-scale flares (M6.6 and X2.2), 
   most parts of the strongly twisted regions remained even after these 
   flares. According to the result in Figure 3(c), because the M1.0 flare 
   occurred outside the strongly twisted region, the strongly twisted region 
   seems not to be deeply related to this flare. In other words, these 
   strongly twisted lines remain in the stable state  despite their 
   continuing growth. On the other hand, in the M1.1 flare, even stronger 
   twisted regions of more than one turn appear and develop. Nevertheless, 
   all the areas dominated by the strongly twisted lines are weakly 
   illuminated by the {\ion{Ca}{2}} image, as shown in Figure \ref{f3}(d), 
   and the most strongly twisted regions remained after this flare, as seen 
   in Figure \ref{f4}. Therefore, this result implies that the strongly 
   twisted lines seem not to be relaxed enough into untwisted lines. These 
   results differ from the common feature of the previous large flares. 
   These topics will be discussed later.
    
   In addition to the above analyses, we show the temporal evolution of the  
   ratio of a fragment of the magnetic flux dominated by values of more than
   half-turn twist to the total flux in Figure \ref{f5}, whose formula is 
   written by 
   \begin{equation}
    F = \frac{\int_{T_n>0.5} B_z dS}{\int B_z dS},
   \end{equation}
   where the d$S$ is a surface element. These fluxes are estimated due to 
   integrate only those $B_z$ in the positive polarities in the same view as 
   in Figure \ref{f4}. From this result, the M6.6 and X2.2 flares on February 
   13 and February 15 also clearly exhibit the store-and-release process of 
   the magnetic helicity during the solar flare, and the flux ratio value in 
   the X2.2 event is about three times larger than that of the M6.6 flare. 
   On the other hand, in the last two flares (M1.0 and M1.1 flares), we can 
   see again that the continuous shear and twist motions of the positive 
   sunspot after the large flares regenerate the strongly twisted lines; 
   consequently, a flux comparable to that before the previous large flare 
   is regained. \cite{2012ApJ...752L...9J} also indicated that the relative 
   magnetic helicity or helicity injection retain their increasing trend 
   after X2.2 flares. Nevertheless the magnetic flux dominated by the strongly 
   twisted lines is mostly unchanged during the M1.0 and M1.1 flares. 

   \subsubsection{Decrement of the Magnetic Twist during the Major Flares}
   We investigate in more detail about the decrement of the magnetic twist 
   during each flare event to determine how much magnetic twist is needed to 
   cause large flares. Figure \ref{f6} shows the distribution map for the 
   values of the magnetic twist (vertical axis) versus the $B_z$ component 
   (horizontal axis). We clearly see that most dotted points in the upper 
   area, above the horizontal dotted line, disappear after each flare; thus, 
   the density of the distribution of less than half-turn twist seems to be 
   partially enhanced. This result suggests that the strongly twisted lines 
   (in excess of half-turn twist) relax into less-twisted lines. Because many 
   dotted points are, even after each flare, widely distributed in the twist 
   range $ T_n < 0.5$, the buildup of twisted field lines having more than 
   half-turn twist is an important process for generating the large flares. 
   Moreover, the dotted points above the horizontal dashed line before the 
   X2.2 flare are obviously greater in number and composed of a more twisted 
   region than those before the M6.6 flare, as shown in Figure \ref{f5}. On 
   the other hand, Figure \ref{f7} shows the distribution maps before and 
   after the M1.0 and M1.1 flares. In this case, most of the dotted points 
   remained in the upper region (values greater than half-turn twist) even 
   after the flare and these distributions seem not to change dramatically 
   during these flares, as we expected from the above results.
  
  \section{Discussion}
   The results shown in Section 3 demonstrated that strong (more than a 
   half-turn) magnetic twist is closely related to produce the large flares, 
   i.e., the X2.2 and M6.6 flares, because most of the strong twist 
   disappeared after these two strong flares. On the other hand, the 
   distribution of the magnetic twist did not change dramatically before and 
   after the M1.0 and M1.1 flare events, in which the  strongly twisted lines 
   ($T_n>0.5$) and even those  with more than one-turn twist ($T_n > 1.0$) 
   were built up before these weak flares. In this section, we investigate 
   the temporal evolution of the twisted field lines and its magnetic topology
   before each flare to explain why the strongly twisted field lines remained 
   after the M1.0 and M1.1 flares. Eventually, we give a suggestion related to 
   the mechanism of the confined and ejective eruptions.

  \subsection{Temporal Evolution of the Twisted Field Lines}
   Figure \ref{f8} shows selected 3D field lines before each flare. The 
  strongly twisted lines (orange) lie on the neutral line in each case. 
  Although some weakly twisted lines (blue) in addition to the strongly 
  twisted lines also form the shear structure at 16:00 UT on February 13, 
  in the last two flares (M1.0 and M1.1 flares), some field lines whose 
  footpoints are rooted in regions of strong magnetic field seem to surround 
  the strongly twisted lines. These relationships between the strongly 
  ($T_n > 0.5$) and weakly ($T_n < 0.5$) twisted field lines, especially 
  those surrounding the strongly twisted lines, might imply some influences 
  on the dynamics of each flare.

   We investigate in detail the relationship between these strongly and 
  weakly twisted lines before each flare by using the connectivity map in 
  the upper panels in Figure \ref{f9}, which focuses on specific field 
  lines for which both footpoints are rooted in the regions of the strong 
  magnetic field ($B_z>$500G and $B_z<-$500G). Consequently, both footpoints 
  of the selected field lines are rooted in the regions shown in white in 
  both polarities, where the white lines correspond to the contour at 
  $|B_z|=$500G. 

  These field lines are shown in the middle panels in Figure \ref{f9}.    
  According to these results, at 16:00 UT on February 13, all the selected 
  closed field lines form a similar shear structure, which are strongly 
  related to M6.6 flares because {\ion{Ca}{2}} image is strongly illuminating 
  in those footpoints in Figure \ref{f3}. Afterward, these twisted lines are 
  deformed due to the subsequently strong sheared and twisted motion of the 
  sunspot. Consequently, at 00:00 UT on February 15, the weakly twisted lines 
  (blue) seem to partially cover the strongly twisted lines (orange). In fact,
  X2.2 flare can be considered to be induced by the strongly twisted lines 
  over the half turn twist from Figure \ref{f3}. Eventually, on February 16, 
  most of the strongly twisted lines were covered by weakly twisted lines. 
  These results clearly show that the weakly twisted lines, which constituted 
  the core field on February 13, play a role in the overlying field lines 
  surrounding the strongly twisted lines. Therefore, as time passes, the 
  strongly twisted lines seem to fall into an unfavorable condition to escape 
  from the lower coronal region. 

  In the lower panels in Figure \ref{f9}, we show a side view of the 3D 
  field lines before each flare in order to more understand the overview 
  profiles of the field lines qualitatively. We clearly see at 00:00 UT on 
  February 15 that the part of the weakly twisted field lines in blue 
  extends to a higher position compared to those on February 13. On the 
  other hand, although these extended loops are still remained at 00:00 
  UT and 07:00 UT on February 16, they seem to become a bit more compact than 
  earlier. This might be related to the report by \cite{2012ApJ...757..149S}, 
  which indicated that this active region decreased in volume after the X2.2 
  flare. 

  Figure \ref{f10} shows the flux ratio of the weakly twisted lines connecting 
  both polarity regions in white in Figure \ref{f9} to the magnetic flux 
  related to the entire closed field, which is composed of the strongly and 
  weakly twisted lines connecting same areas. All of the flux values are 
  estimated due to integrate only those $B_z$ in the positive polarities 
  before the flares on February 15 and 16, and the flux ratio formula is 
  defined as 
  \begin{equation}
   F_{r} = \frac{\int_{0<T_n<0.5} B_z dS}{\int B_z dS}, 
  \end{equation}
  where d$S$ is a surface element. Note that because we are interested in 
  the deformed weakly twisted lines due to subsequently strong twisted and 
  sheared motion of the sunspot after M6.6 flare,  the value on 
  February 13 is not plotted in Figure \ref{f10}. 

  This result shows us that this flux ratio remains 62$\%$-67$\%$ during 
  the three flares. Therefore, these results reveal no large variations in 
  the magnetic flux. Before the X2.2 flare on February 15, even though the 
  weakly twisted lines whose footpoints are both at $|B_z|>$500 G had already 
  deformed, they covered the strongly twisted fields only in part. This 
  surrounding condition might allow the escape of the strongly twisted field 
  from the lower coronal region. On the other hand, the magnetic field 
  configuration is changed by the strong shear and twist motions of the 
  sunspot; thus the weakly twisted field lines covered the entire strongly 
  twisted field region on February 16. Furthermore, along with the results 
  in Figure \ref{f10}, because the 'total' magnetic energy accumulated in 
  the root of the the weakly twisted lines is superior to that of the strongly 
  twisted field lines, these probably have the potential to confine the 
  dynamics of the strongly twisted field in the M1.0 and M1.1 flares. For 
  example, this confinement indicates the suppression of the exhaustion at 
  the magnetic reconnection site and consequent rapid energy release.

   \subsection{Overlying Field Lines}
   In previous subsection, we discuss the magnetic structure inducing the 
   X- and M-class solar flares in terms of the temporal evolution and 
   topologies of the twisted field lines formed in the lower corona. However, 
   the overlying field lines surrounding these twisted lines also play an 
   important role in determining whether CMEs are produced or not even if 
   the twist value in the twisted line is more than one-turn twist 
   (\citealt{2010ApJ...725L..38G}).  Therefore, we need to discuss the 
   overlying field lines quantitatively. The decay index is one of 
   quantitative indexes of the overlying fields by estimating how rapidly 
   the strength of the magnetic field decreases with height, which is 
   given by the following equation, 
   \[n(z) = - \frac{z}{|\vec{B}|}\frac{\partial |\vec{B}|}{\partial z}.\]
   The decay index is often referred to an estimation of a criterion for 
   the torus instability whose threshold lies in the range of 
   1.1$\leq n_c \leq$2, which was  introduced by \cite{2006PhRvL..96y5002K} 
   and \cite{2010ApJ...718.1388D}. An eruption of the flux rope was 
   numerically confirmed  by \cite{2007AN....328..743T}, 
   \cite{2010ApJ...719..728F} and \cite{2010ApJ...708..314A}. 

   In this study, as mentioned in Section 2, we can discuss only the 
   closed field lines in the central area within a height of around few dozens
   of Mm due to the problem associated with lateral boundary conditions. 
   Because of this, decay index in the overlying field lines higher than few 
   dozens of Mm is strongly affected by these boundary conditions. Furthermore
   because decay index cannot be applied to the non-potential field structure 
   retrieved with our mode within a height of around few dozens of Mm, it is 
   difficult to discuss it in the framework of our study. However, 
   \cite{2012ApJ...748L...6N} already calculated the decay index related to 
   the NLFFF (\citealt{2012SoPh..tmp...67W}) in AR11158 in long term period on
   February 11-16. They showed the long-term temporal evolution of the decay 
   index at each height and concluded that the onset of eruptions does not 
   depend critically on the long-term evolution of the decay index of the 
   background field before CMEs. In other words, this result might support 
   that the temporal evolution of the twisted field lines play an important 
   role in controlling the flares and CMEs in this active region.

   In fact, \cite{2011AGUFMSH13B1965Y} reported that partial Halo and Halo 
   CMEs were observed associated with M6.6 and X2.2 flares, on the other hand,
   we were not able to observe the CMEs just after M1.0 and M1.1 flares (http://cdaw.gsfc.nasa.gov/CMElist/). Only one CME originated from AR11158 was 
   observed  after X2.2-class flare occurred. However, the flare which is 
   source of this CME occurred in the edge of this active region. Therefore 
   these observations might support that flux rope eruption is confined by 
   the weak twisted line plotted in blue in Figure \ref{f9} during M1.0 and 
   M1.1 flares.

  \subsection{Formation of the more strongly twisted Flux Tube before 
              an Eruption}
   In this study, the strongly twisted field lines before M6.6 and X2.2
   flares distribute in $0.5 < T_n< 1.0$, which implies that the 
   magnetic configuration is stable against the MHD instability. Although, 
   these magnetic twists obtained from our study seem to be weak to induce 
   the large flares, this is because the NLFFF approximation omits the 
   necessary physics in a dynamic process. i.e., missing the tether-cutting 
   process. Some authors also have supported tether-cutting reconnection as a 
   feasible process causing eruption in this active region as described in  
   Section 1. Because of conservation of the magnetic helicity, tether-cutting
   reconnection in the strongly twisted lines generates longer and more 
   strongly twisted field lines just before an eruption. For example, 
   \cite{2010ApJ...717L..26A} and \cite{2011ApJ...742L..27A} showed the 
   strongly twisted flux tube along the PIL through the tether-curring 
   reconnection in the twisted field lines due to the flux cancellation 
   process or converging flows on the photosphere. Eventually, this flux 
   tube is successfully launched from the lower corona. More recently, 
   \cite{2012ApJ...760...31K} indicated two types of emerging flux 
   that produce the long strongly twisted field lines in the pre-existing 
   coronal magnetic field through tether-cutting-like reconnection, where 
   the initial pre-existing field is assumed to form uniformly sheared 
   arcades in the linear force-free approximation. They reported that the 
   angle between the emerging flux and the pre-existing shear field lines 
   is important. Thus, tether-cutting process would be feasible of producing 
   large flares even the accumulated twist in the solar active region is 
   less than one-turn twist. 
 
  \section{Summary}
\begin{figure}
  \epsscale{1.}
  \plotone{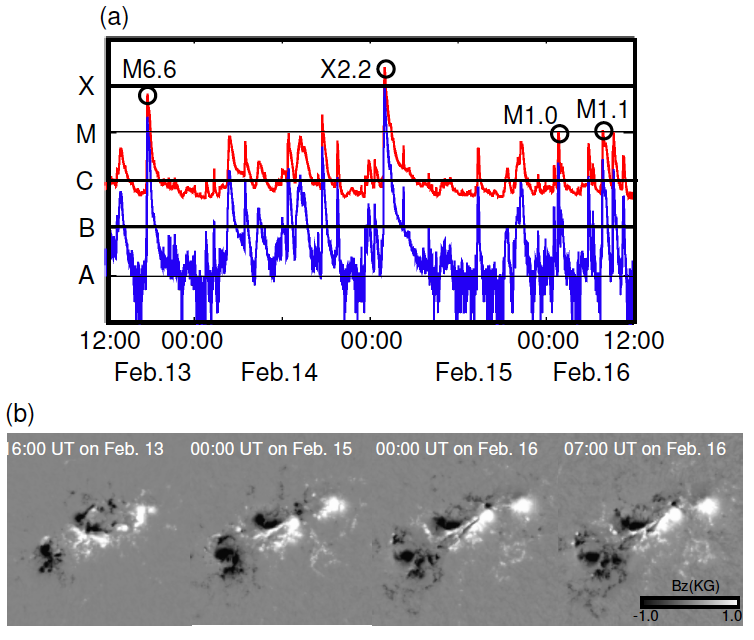}
  \caption{
           (a) Time profile of X-ray flux measured by GOES satellite on 2011 
               February 13-16. The solar X-ray outputs in the 1.0-8.0 
               {\AA}(red) and 0.5-4.0 {\AA}(blue) passband are plotted. The 
               four flares (M6.6, X2.2, M1.0, and M1.1) marked by circles are 
               analyzed in this study. 
           (b) Vector fields 40-90 min before each flare. The distributions of 
               the normal component of the magnetic field($B_z$) at each time 
               are plotted in gray scale. 
          }
  \label{f1}
  \end{figure}

\begin{figure}
  \epsscale{.85}
  \plotone{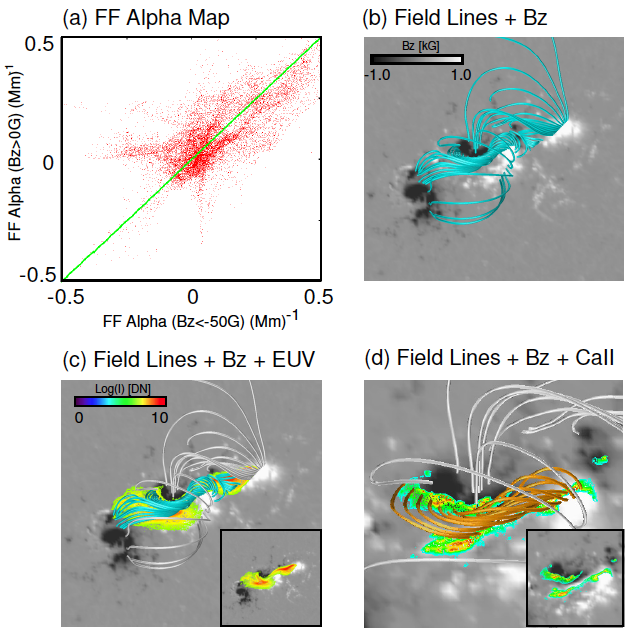}
  \caption{
           (a) Distribution of force-free $\alpha$ map of reconstructed field 
               at 00:00 UT on February 15. The closed field lines are focused 
               and estimated in the range of about 74 $\times$ 74 ($Mm^2$) in 
               the central area. Vertical and horizontal axes represent the 
               values of the force-free $\alpha$ in opposite footpoints on 
               each field line. These values are estimated in the plane at 
               1688 km above the photosphere, and the field lines are traced 
               from the region in which the values of the magnetic flux are 
               less than $-$50 G. Green line indicates the function of $y=x$. 
           (b) Selected field lines in blue extrapolated from vector field 
               from HMI/{\it SDO} observed at 00:00 UT on February 15 plotted 
               over the distribution of the $B_z$ component. The field lines 
               are traced from positive polarity values greater than 250 G.
           (c) EUV images in 94 {\AA} from AIA/{\it SDO} observed at 
               23:59:28 UT on February 14; features whose intensity is more 
               than 1.0$\times 10^{5}$(DN) are plotted over (b). Blue field 
               lines capture the region in which the EUV images are strongly 
               enhanced. Others are plotted in gray. Inset shows the same 
               figure without field lines.
           (d) Another set of field lines plotted over the $B_z$ distribution 
               at the same time as (b)-(c) and {\ion{Ca}{2}} image from 
               SOT/{\it Hinode} observed at 01:50:18 UT on February 15. 
               Footpoints of orange field lines are rooted in the region where
               the {\ion{Ca}{2}} image is strongly enhanced (values greater 
               than 1000(DN)). Inset shows the same figure without field 
               lines. 
          }
  \label{f2}
  \end{figure}

\begin{figure}
  \epsscale{1.}
  \plotone{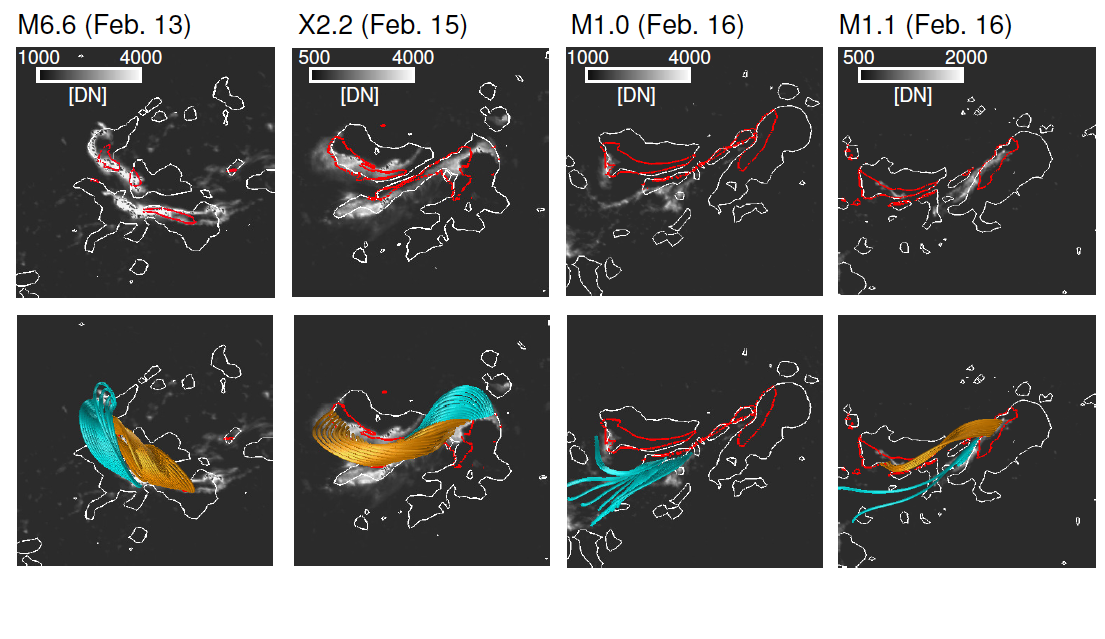}
  \caption{
           Upper panels show the normal component of magnetic field and 
           twist profile plotted on the {\ion{Ca}{2}} image before the each 
           flare. White lines represent the contours of normal component of 
           magnetic field ($|B_z|$=625 G) observed at same times shown 
           in Figure \ref{f1}(b). Red lines show the magnetic twist 
           ($T_n=0.5$) obtained from the NLFFF extrapolated from each vector 
           field. Regions surrounded by red lines are occupied by strongly 
           twisted lines ($T_n > 0.5$). Gray scale shows {\ion{Ca}{2}} image 
           observed at 17:35:38 UT on February 13, 01:50:18 UT on 
           February 15, 01:40:39 UT and 07:42:13 UT on February 16, 
           respectively. Lower panels show the selected magnetic field lines 
           traced from the regions in which {\ion{Ca}{2}} illuminates 
           strongly. Orange and blue field lines represent twist values more 
           and less than half-turn twist ($T_n=0.5$), respectively.
          }
  \label{f3}
  \end{figure}

\begin{figure}
  \epsscale{1.}
  \plotone{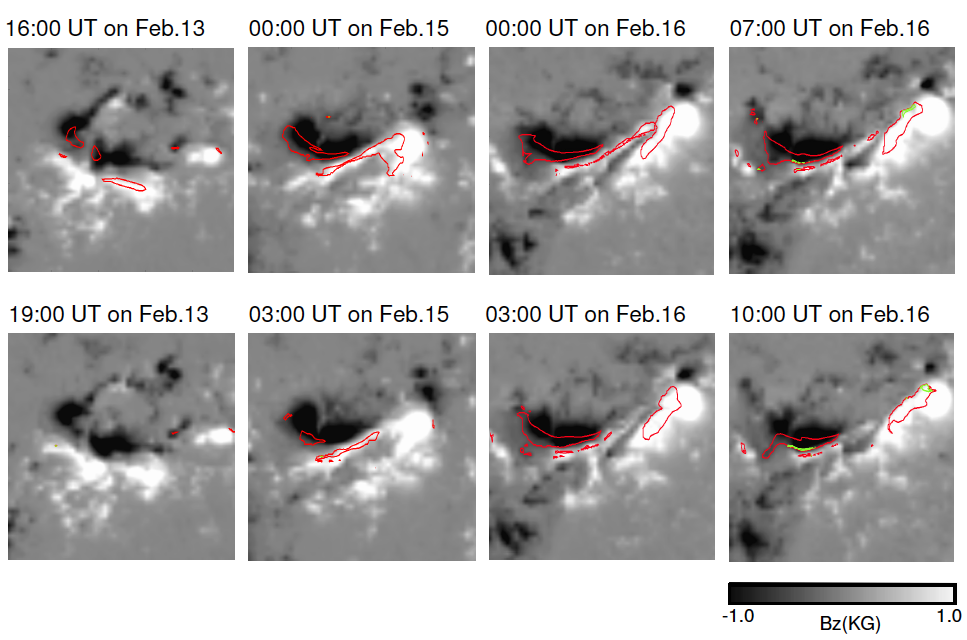}
  \caption{
           Temporal evolution of the magnetic twist with the distribution of 
           $B_z$ component in gray scale, which corresponds to central area
           of the active region. Upper and lower panels represent 40-90 min 
           before and after each flare (M6.6, X2.2. M1.0, and M1.1) 
           respectively. Red and green lines represent the contours of 
           magnetic twist $T_n = 0.5$ and $T_n=1.0$, respectively. Regions 
           surrounded by red and green lines indicate strongly twisted 
           regions of $T_n > 0.5$, and $T_n > 1.0$, respectively. 
          }
    \label{f4}
    \end{figure}

\begin{figure}
  \epsscale{1}
  \plotone{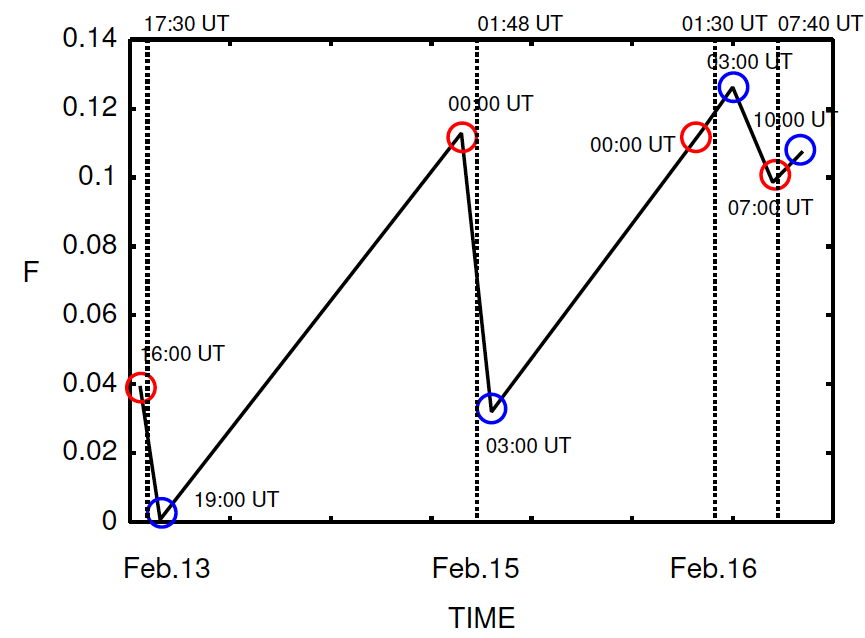}
  \caption{
           Temporal evolution of the ratio of the fragment of magnetic flux 
           dominated by values of more than half-turn twist to the total 
           magnetic flux from 16:00 UT on February 13 to 10:00 UT on February 
           16. All flux is estimated in the positive polarity. Vertical 
           dashed lines indicate occurrence of each flare. Red and blue 
           circles indicate time at which each NLFFF was reconstructed 
           before and after each flare, respectively.
          }
   \label{f5}
   \end{figure}

\begin{figure}
  \epsscale{1.}
  \plotone{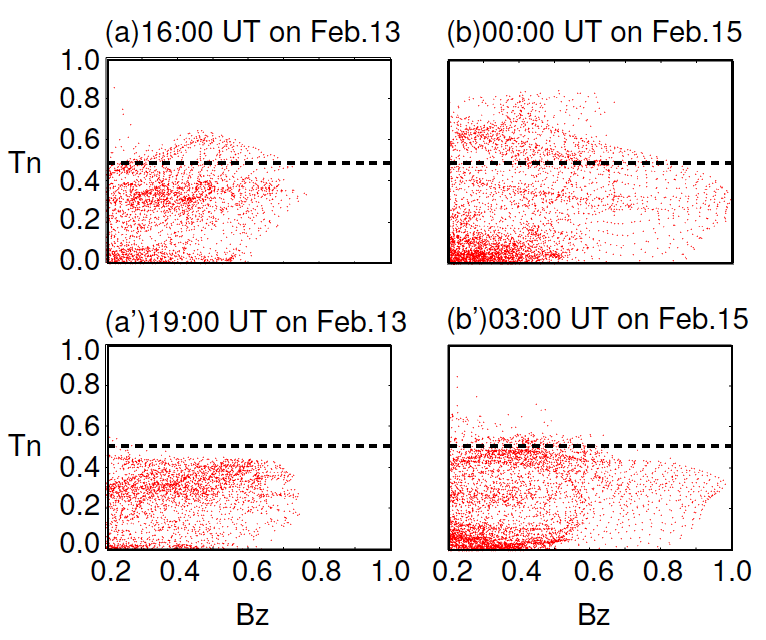}
  \caption{
          (a),(b) Distribution map related to the twist (vertical axis) and 
                  $B_z$ components (horizontal axis) at 16:00 UT on February 
                  13 and 00:00 UT on February 15 (before M6.6 and X2.2 flares,
                  respectively). $B_z$ component is focused on values in 
                  excess of 500G, whose normalized value corresponds to 0.2. 
                  Horizontal dashed line indicates the value of half-turn 
                  twist($T_n = 0.5$).  
       (a'), (b') Maps in same format at 19:00 UT on February 13 and 03:00 UT 
                  on February 15 corresponding to the period after the M6.6 
                  and X2.2 flares, respectively.                  
          }
  \label{f6}
  \end{figure}

\begin{figure}
  \epsscale{1.}
  \plotone{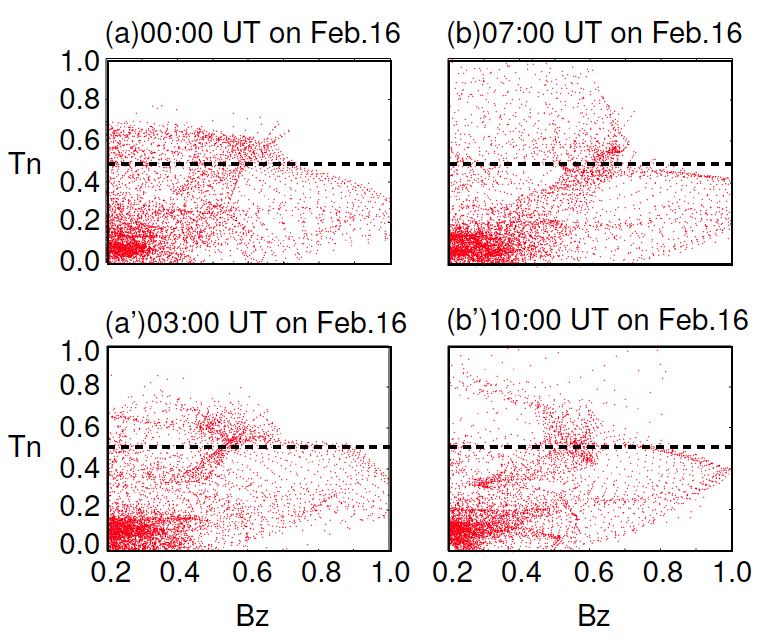}
  \caption{
           Distribution maps showing magnetic twist value (vertical axis) 
           versus $B_z$ component (horizontal axis) before and after M1.0 
           and M1.1 flares; formats are the same as in Figure \ref{f6}. 
           }
  \label{f7}
  \end{figure}

\begin{figure}
  \epsscale{1}
  \plotone{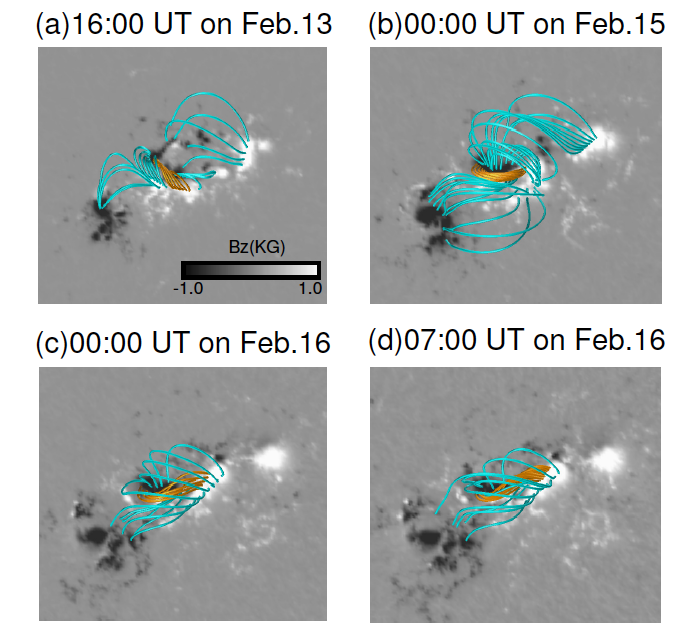}
  \caption{
           Selected field lines plotted over $B_z$ component 40-90 min 
           before each flare. Orange represents twist values of $T_n > 0.5$; 
           blue corresponds to $T_n < 0.5$. Both plotted from positive 
           polarities of greater than 250G.
          }
   \label{f8}
   \end{figure}

\begin{figure}
  \epsscale{1.}
  \plotone{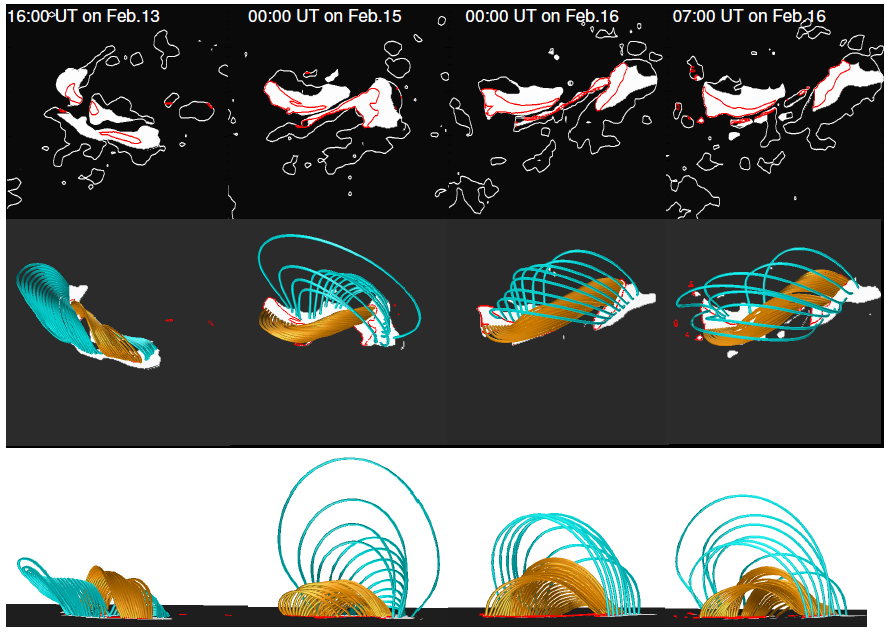}
  \caption{
           Top panels show the connectivity maps in black and white scale are 
           represented at 40-90 min before each flare. White lines represent 
           the contours of $|B_z|=500$ G; red line indicates contour of 
           magnetic twist corresponding to $T_n$=0.5. White regions are 
           specific areas in which the footpoints of the closed field lines 
           connecting the regions surrounded by white contours are rooted. 
           In the middle panels, the field lines are plotted over the upper 
           panels except for the white lines. All of their footpoints are 
           rooted in the regions in the white areas. Orange and blue field 
           lines have magnetic twist values greater and less than $T_n$=$0.5$,
            respectively. In the lower panels, side views of 3D field lines. 
           }
  \label{f9}
  \end{figure}

\begin{figure}
  \epsscale{1}
  \plotone{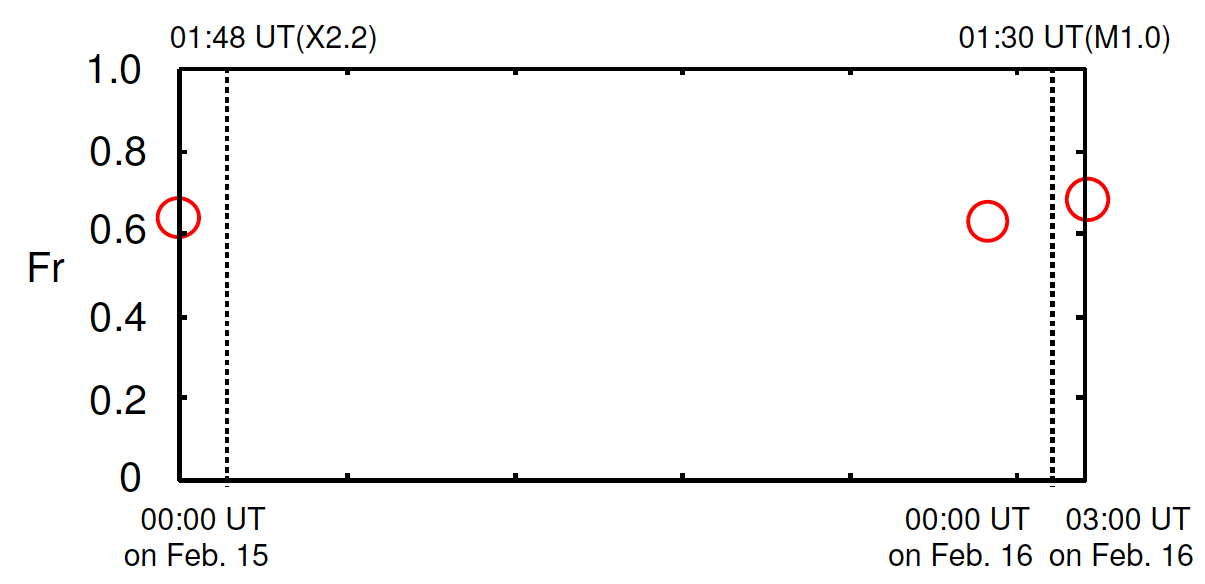}
  \caption{
           Temporal evolution of the ratio of the fragment of the magnetic
           flux less than $T_n =$0.5 to the total integrated flux. In this
           case, all of the magnetic flux is estimated by the closed field 
           lines connecting both polarities in white in Figure \ref{f9} and 
           integrated only those $B_z$ in the positive polarities.
          }
   \label{f10}
   \end{figure}

   This paper presented the 3D magnetic structures of AR11158, which produced 
   a X-class and several M-class flares on 2011 February 13-16. We focus 
   on four flares, M6.6, X2.2, M1.0, and M1.1, as shown Figure\ref{f1}(a), 
   which are analyzed in terms of the magnetic twist obtained from the NLFFF 
   and its variation before and after each flare. These NLFFFs were obtained 
   from the MHD relaxation method developed by 
   \cite{2011ApJ...738..161I}, 
   \cite{2012ApJ...747...65I}, and 
   \cite{2012ApJ...760...17I}. 
   Our previous studies were focused on a X-class flare where we discussed 
   its magnetic structure and physical condition in terms of the magnetic 
   twist in the solar active region 10930. However, in this present study, 
   we are able to compare them quantitatively in the various class-flares 
   occurred in the active region 11158 and eventually gave a suggestion 
   related to the ejective and confined eruptions of CME as well as the 
   occurrence conditions of X- and M-class flares.
    
   First of all, we compared the magnetic twist obtained from the 3D field 
   lines before each flare event with {\ion{Ca}{2}} images obtained from 
   SOT/{\it Hinode}. Particularly in the M6.6 and X2.2 flares, we found that 
   the footpoints of the strongly twisted field lines whose values are larger 
   than the half-turn twist (i.e., $T_n>0.5$) corresponded to the locations 
   well within the strong enhancement of {\ion{Ca}{2}}, which is consistent 
   with our previous study (\citealt{2011ApJ...738..161I}). These results 
   show that dramatic magnetic reconnection occurred in the strongly twisted 
   lines ($T_n > 0.5$) in these flares. Furthermore, we found that the 
   magnetic flux ratio of the strongly twisted lines to the total flux before 
   the X2.2 flare was about three times larger than that in the M6.6 flare. 
   On the other hand, magnetic twists larger than $T_n = 1.0$ have never 
   seen in either cases which indicates that the magnetic configuration 
   is stable against the ideal MHD instability. The magnetic twists obtained 
   in this study seem a little weak to produce a large flare but this is due 
   to the limitation of the NLFFF omitting some necessary physics in a 
   dynamics, e.g., tether cutting process is one of them. For instance, 
   although the value of $T_n$=0.5 is low for a single flux tube, magnetic 
   reconnection between these twisted lines could produce highly twisted lines
   having more than a one-turn twist ($T_n$ = 1.0) at least; therefore, active
   regions accumulating the twisted field lines at 0.5$< T_n <$1.0 could be 
   capable of producing large flares.

   A comparison of the conditions before and after each flare revealed that 
   strongly twisted lines were built up before each flare; they disappeared 
   after the M6.6 and X2.2 flares, whereas the weakly twisted lines 
   ($T_n < 0.5$) remained. On the other hand, although the strongly twisted 
   lines were also built up before the last two flares on February 16, whose 
   magnetic flux strength was comparable to that of the X2.2 flares, the 
   overall distributions of the magnetic twist did not change dramatically 
   after the flare happening. We carefully investigated the temporal evolution
   of the twisted lines and their magnetic topologies before these flares. 
   We found that the twisted field lines are deformed due to the strongly 
   sheared and twisted motion of the sunspot, consequently the weakly twisted 
   lines whose footpoints are at $|B_z|>$500 G covered the strongly twisted 
   lines before M1.0 and M1.1  flares. Because the CMEs were not observed 
   associated with these flares and temporal evolution of the decay index in 
   the overlying field lines surrounding the twisted field lines were not 
   critically depended on an initiation of eruptions, reported by 
   \cite{2012ApJ...748L...6N}, these weakly twisted field lines might confine 
   the activities of the strongly twisted lines even though the magnetic flux 
   of the strongly twisted lines is stronger than that of the lines producing 
   the M6.6 flare and comparable to that of the lines producing the X2.2 
   flare. 
 
   This case is one example of many active regions. Therefore, further 
   analysis of the various solar active regions using observational and 
   numerical approaches is needed to reach on a possible conclusion for 
   flare dynamics. HMI/{\it SDO} successfully observed the vector field 
   of AR 11158 with unprecedented spatial and temporal resolution  and 
   will provide that data for many active regions. We can expect that 
   the results obtained from these data analyses will enable us to better 
   understand the dynamics of this magnetic activities.  

  \acknowledgments
   We are grateful to Drs.\ Vinay Shankar Pandey, Seiji Yashiro, and Tetsuya 
   T Yamamoto for their useful comments. Many thanks to the anonymous referee 
   for carefully checking this paper and constructive comments. S. I. was 
   supported by the International Scholarship of Kyung Hee University. 
   This study was supported by the WCU (World Class University) program 
   (R31-10016) and Korea Meteorological Administration through National 
   Meteorological Satellite Center as well as Basic Science Research 
   Program (2010-0009258; PI: T. Magara) through the National Research
   Foundation of Korea. G.S.C. was supported by the Korea Research Foundation 
   grant funded by the Korean Government (KRF-2007-313-C00324). D.S. was 
   supported by the Grant-in-Aid for Scientific Research (B)''Understanding 
   and Prediction of Trigger of Solar Flares''(23340045, Head Investigator: 
   K.\ Kusano) from the Ministry of Education, Science, Sports, Technology, 
   and Culture of Japan. The computing, data analysis, and visualization were 
   performed using the OneSpaceNet in the NICT Science Cloud. We sincerely 
   grateful to NASA/SDO and the HMI and AIA science team. Hinode is a Japanese
   mission developed and launched by ISAS/JAXA, with NAOJ as domestic partner 
   and NASA and STFC (UK) as international partners. It is operated by these 
   agencies in co-operation with ESA an NSC (Norway).

\clearpage

  \clearpage

  \clearpage

  \clearpage

    \clearpage

   \clearpage

  \clearpage

  \clearpage

   \clearpage

  \clearpage

   \clearpage

  \begin{table}
  \begin{center}
  \caption{Parameters and values of $D=\int |\vec{\nabla}\cdot\vec{B}|^2$dV 
           related to the selected field.  The period is from 2011 February 
           13 to 16.
  \label{tbl-1}}
  \begin{tabular}{crrr}
  \tableline\tableline
  Time  &$\eta_{0}$ & $v_{max}$ & D  \\ 
  \tableline
  16:00 UT on Feb.13  & $2.5   \times 10^{-5}$ & $2.5   \times 10^{-3}$ 
                      & $2.25  \times 10^{-7}$    \\
  19:00 UT on Feb.13  & $2.5   \times 10^{-5}$ & $2.5   \times 10^{-3}$  
                      & $2.41  \times 10^{-7}$    \\
 00:00 UT  on Feb.15  & $2.5   \times 10^{-5}$ & $5.0   \times 10^{-3}$ 
                      & $3.186 \times 10^{-7}$   \\
  03:00 UT on Feb.15  & $2.5   \times 10^{-5}$ & $2.5   \times 10^{-3}$ 
                      & $2.545 \times 10^{-7}$   \\
  00:00 UT on Feb.16  & $5.0   \times 10^{-5}$ & $2.5   \times 10^{-3}$      
                      & $8.518 \times 10^{-7}$   \\
  03:00 UT on Feb.16  & $5.0   \times 10^{-5}$ & $2.5   \times 10^{-3}$  
                      & $1.007 \times 10^{-6}$   \\
  07:00 UT on Feb.16  & $5.0   \times 10^{-5}$ & $5.0   \times 10^{-3}$ 
                      & $7.740 \times 10^{-7}$  \\
  10:00 UT on Feb.16  & $5.0   \times 10^{-5}$ & $2.5   \times 10^{-3}$  
                      & $9.676 \times 10^{-7}$   \\
  \tableline
  \end{tabular}
  \end{center}
  \end{table}




\end{document}